\documentclass[%
 reprint,
 amsmath,amssymb,
 aps,
 prl,
floatfix,
]{revtex4-2}
\usepackage{graphicx}
\usepackage{dcolumn}
\usepackage{bm}

\usepackage{tabularx}
\usepackage[caption=false, justification=centerlast]{subfig}

\usepackage{xcolor}

\begin{document}


\title{
Synchronization in the presence of time delays and inertia: Stability criteria
}

\author{Dimitrios Prousalis}%
 \email{dprou@pks.mpg.de}
\author{Lucas Wetzel}
 \email{lwetzel@pks.mpg.de}
\affiliation{%
Max Planck Institute for the Physics of Complex Systems, Dresden, Germany
}%



\date{\today}

\begin{abstract}

Linear stability of synchronized states in networks of delay-coupled oscillators depends on the type of interaction, the network and oscillator properties. 
For inert oscillator response, found ubiquitously from biology to engineering, states with time-dependent frequencies can arise.
These generate side bands in the frequency spectrum or lead to chaotic dynamics.
Stability analysis is difficult due to delay-induced multistability and has only been available via numerical approaches.
We derive criteria and conditions
that enable fast and robust analytical linear stability analysis
based on the system parameters.
These apply to arbitrary network topologies, identical oscillators and delays.
\end{abstract}
\maketitle



\textit{Introduction:}---Self-organized synchronization can be observed in chemical oscillators, embryonic development, circadian clocks, ranging to power grids and the orchestration of mobile communications, microelectronic and mechanical systems \cite{Buck1988, Trees2005, Schoell2010, Oates2012, Rohden2012, Motter2013, Dewenter2015, Rodrigues2016, Kuznetsov2017, Koskin2018}.
This type of synchronization has been considered for electronic networks 
since the $1980$'s due to its robustness and as its properties scale advantageously with growing system size  \cite{Lindsey1985, Santini2009}.
In application however, it did not prevail over hierarchical synchronization as the necessary theoretical framework to guide architecture design was not available \cite{Lindsey1985}.
Within phase oscillator models the dynamics in networks of coupled oscillators can be studied \cite{Kuramoto1984, Joerg2015}.
This includes the effects of inevitable time delays in the coupling.
These lead to phenomena like multistability of synchronized states \cite{Schuster1989}.  
Another aspect of the oscillators' dynamics has recently come into focus, inert response to external stimuli.
Examples are the inertia of mechanical oscillators, signal filtering in electronics, or biochemical transport and conversion processes in cellular oscillators \cite{Gupta2014, Olmi2014}.
Inert system response in $2^\textrm{nd}$ order phase models can trigger bifurcations of synchronized states with constant frequency \cite{Diekmann1995, Pollakis2014}.
Frequency modulation occurs, side bands arise in the spectrum and synchronized states with constant phase relations become unstable.

In this work we derive stability criteria for in- and anti-phase synchronized states in networks of delay-coupled oscillators with inertia. 
These depend only on the physical properties of the oscillators and the network and can guide, e.g., the architecture design of synchronization layers in networks of mutually coupled electronic oscillators.
In parameter space plots we then discuss the linear stability of in- and anti-phase synchronized states in general.
Our criteria simplify studying the physical properties of synchronization over large parameter regimes and for, e.g., large delays and large number of oscillators.
We then discuss how linear stability depends on physical properties such as time delay, inertia, damping or dissipation, interaction strength and network topology.
These generic concepts can then be related to application specific concepts like, e.g, the loop gain and bandwidth of electronic oscillators or the dissipation coefficients in power grids \cite{Filatrella2008, Schaefer2015, Gorjao2020, Wetzel2021}. 
Additionally, we present a condition connecting these quantities.
If fulfilled, linear stability is guaranteed and hence no bifurcations occur.

\textit{Networks of delay-coupled oscillators with inertia:}---
The dynamics in such networks can be studied within the following set of coupled delay-differential equations
\begin{equation}
 m\,\ddot{\theta}_k(t)+\gamma\dot{\theta}_k(t)=\omega+\frac{K}{n_k}\sum_{l=0}^N c_{kl}h\left(\Delta\theta_{kl}(t,\,\tau,\,v)\right),
  \label{eq:model_2nd}    
\end{equation}
where $\omega\,\in\,\mathbb{R}$ denotes the intrinsic frequency,
$h(\cdot)$ the coupling function,
$K\geq0\,\in\,\mathbb{R}$ the coupling strength,
$m\geq0\,\in\,\mathbb{R}$ an inertial parameter,
$\gamma>0\,\in\,\mathbb{R}$ a damping parameter,
$n_k\geq0\,\in\,\mathbb{N}_0$ the number of inputs of oscillator $k$,
$\theta_i(t)\,\in\,\mathbb{R}$ for $i=\{k,l\}$ the phases of the oscillators' output signals with $\dot{\theta}$ and $\ddot{\theta}$ denoting their first and second time derivatives,
$c_{kl}$ the components of the network's adjacency matrix, being either $1$ if there is a connection from oscillator $l$ to $k$, or $0$ otherwise.
$\Delta\theta_{kl}(t,\,\tau,\,v)=(\theta_{l}(t-\tau)-\theta_{k}(t))/v$ is the phase-difference between $k$ and an input $l$.
Here $v\,\in\,\mathbb{N}$ denotes the division of the instantaneous output frequency of the oscillators, e.g., induced by a frequency divider.
This is well known from, e.g., periodic cross-coupling signals in networks of electronic oscillators \cite{Best2003}.
Note that Eqs.~(\ref{eq:model_2nd}) reduce to the classical first order Kuramoto model for sinusoidal coupling $h(\cdot)=\sin(\cdot)$, zero coupling delay $\tau=0$, damping coefficent $\gamma=1$, and inertia $m=0$.
We study in- and anti-phase synchronized states making the Ansatz
\begin{equation}
 \theta_k(t)=\Omega t +\beta_k + \epsilon q_k(t),
 \label{eq:ansatz}
\end{equation}
where $\Omega$ denotes the frequency of a synchronized state, $\epsilon q_k(t)$ a small perturbation ($\epsilon\ll 1$), and $\beta_k$ a phase-offset.
The properties of synchronized states can then be obtained in $\mathcal{O}(\epsilon^0)$ by using the Ansatz (\ref{eq:ansatz}) and Eqs.~(\ref{eq:model_2nd})
\begin{equation}
 \gamma\,\Omega=\omega+K\,h\left(-\frac{\Omega\tau+\beta}{v}\right),
 \label{eq:SyncStates}
\end{equation}
where $\beta=\beta_l-\beta_k$ equals to $0$ or $\pi$ for in- and anti-phase synchronized states in networks of identical oscillators, respectively.
From $\mathcal{O}(\epsilon^1)$ the dynamics of small perturbations in the Laplace domain can be inferred:
\begin{equation}
 e^{\lambda\tau}\left(m\lambda^2+\gamma\lambda+\alpha\right)q_k(\lambda) = \alpha\sum\limits_{l=1}^N\,d_{kl}\,q_l(\lambda),
 \label{eq:PerturbationDynamicsLaplace}
\end{equation}
\onecolumngrid

\begin{figure}[pt!]
\includegraphics[width=7in]{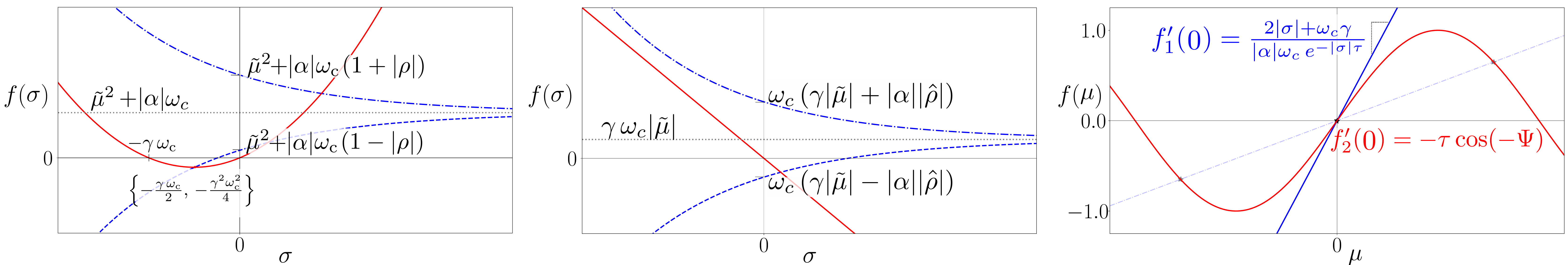}
 \caption{Graphical solutions to Eqs.~(\ref{eq:real_and_imaginary_parts}). Shown for $\alpha=-|\alpha|<0$ (left), $\alpha>0$, $\tilde{\mu}<0$ (middle), $\sigma\geq 0$, when $\tilde{\mu}=0$ (right).}
 \label{fig:graphic_solutions}
\end{figure}

\twocolumngrid
\noindent
where $\alpha=\frac{K}{v} \, h'(\frac{-\Omega \tau+\beta}{v})$ denotes a steady state parameter, $d_{kl}=c_{kl}/n_k$ the components of the normalized adjacency matrix $\mathbb{D}$, and $\lambda= \sigma+i\mu$ the complex frequency.
Rewriting in matrix form, we identify the eigenvalue problem $\zeta\,\vec{q}=\mathbb{D}\,\vec{q}$ and the characteristic equation 
\begin{equation}
  \lambda^2+ \omega_c \gamma \lambda + \alpha \omega_c (1- \zeta e^{-\lambda \tau})=0\, ,
  \label{eq:characteristic_equation}
\end{equation} 
where $\omega_c=m^{-1}$ and $\zeta=|\zeta|e^{i \Psi}$ are the eigenvalues of the normalized adjacency matrix $\mathbb{D}$. 
%
These $\zeta$ relate to the perturbation modes in the network and each generates an infinite discrete set $\Lambda_\zeta$ of solutions $\lambda$.
For diagonalizable $\mathbb{D}$ arbitrary perturbations can be expressed by linear combinations of $\vec{q}$, the eigenvectors.
The eigenvector that induces a global shift of all phases $\vec{q}=(1,\,1,\,\dots,\,1)$ has the eigenvalue $\zeta=1$ since $\sum_l\,d_{kl}=1$.
It does not affect the synchrony of the system and will be excluded in the following discussions.
From dynamical systems theory it is known that the largest $\sigma$ in the union $\cup_{\zeta\neq 1}\Lambda_\zeta$ dominates the long term dynamics of the perturbations.
If the largest $\sigma>0$ then perturbations grow and the state is linearly unstable.
If all $\sigma<0$ the system is linearly stable.
From Eq.~(\ref{eq:characteristic_equation}) we find that if $\alpha = 0$ then $\lambda_1=0$ and $\lambda_2=-\omega_c \gamma$.
Hence, $\alpha=0$ relates to marginally stable solutions and will also not be considered in the following. 

\textit{Derivation of stability criteria:}---
For first order Kuramoto models with time delays, i.e. without inertia, Earl and Strogatz derived a criterion that determines linear stability of synchronized states \cite{Earl2003}.
It concludes that synchronized states in networks of delay-coupled oscillators with arbitrary coupling topology are linearly stable if and only if $\alpha=K\,h'\left(-\Omega\tau\right)>0$. 
For Kuramoto models with time delay and inertia it has been shown that this criterion cannot sufficiently predict linear stability \cite{Wetzel2017}.
There is no known closed form solution to second order exponential polynomials like Eq.~(\ref{eq:characteristic_equation}).
Such solutions can be obtained numerically but require a careful choice of initial conditions and become increasingly difficult for large time delay and network size.
In previous works, conditions that connect inertial properties with the interaction strengths and properties of the synchronized states that prevent instability have been found \cite{Lindsey1985, Schaefer2015, Dai2018}.

Here, we introduce stability criteria that allow to predict linear stability of in- and anti-phase synchronized states in networks of delay-coupled oscillators with inertia for any set of parameters.
Furthermore, we extend the previously found conditions and connect them to properties of the topology \cite{Schaefer2015}.
With $\lambda= \sigma+i\mu$ in Eq.~(\ref{eq:characteristic_equation}) and separate the real and imaginary parts:
\begin{subequations}
\begin{eqnarray}
  \sigma^2+ \omega_c \gamma \,\sigma &=& -\alpha\,\omega_c\left(1- |\zeta|\cos(\mu\tau-\Psi)\,e^{-\sigma\tau}\right)+\mu^2, \qquad \label{eq:real_part}\\[5pt]
  2\sigma\mu &=&\, -\omega_c\left(\mu\,\gamma+\alpha|\zeta|\sin(\mu\tau-\Psi)\,e^{-\sigma\tau}\right). \qquad \label{eq:imaginary_part}
\end{eqnarray} 
\label{eq:real_and_imaginary_parts}
\end{subequations}
Squaring and adding these equations we obtain
\begin{equation}
  (\sigma^2-\mu^2+\omega_c\gamma\sigma+\alpha\omega_c)^2+(2\sigma\mu+\omega_c\gamma\mu)^2=(\alpha\omega_c|\zeta|)^2e^{-2\sigma\tau}.
  \label{eq:Squared_Real_Imag_Parts}
\end{equation}

We begin by addressing one direction of the known stability criterion presented in \cite{Earl2003}.
%
For second order phase models we show that \textit{if $\alpha < 0$, there always exists at least one $\sigma > 0$} and hence the states in Eq.~(\ref{eq:SyncStates}) 
are unstable. 
Setting $\alpha=-|\alpha|$ and $\rho=|\zeta|\cos(\mu\tau-\Psi)$ in Eq.~(\ref{eq:real_part}) we find after rearranging
\begin{equation}
  \sigma^2+ \omega_c \gamma \sigma = |\alpha|\,\omega_c\left(1- \rho e^{-\sigma\tau}\right)+\mu^2,
  \label{eq:proof1}
\end{equation} 
where $\rho\,\in\,[-1,\,1]$, since $|\zeta|\leq 1$ as can be shown from Gershgorin's circle theorem \cite{Gerschgorin31}, see Supplementary Material.
Using the boundedness of the $|\zeta|$'s and Eq.~(\ref{eq:proof1}) we prove the proposition graphically, see Fig.~\ref{fig:graphic_solutions} (left).
%
The left hand side (l.h.s.) of Eq.~(\ref{eq:proof1}) is quadratic in $\sigma$ and crosses the x-axis at the origin $\sigma_1=0$ and at $\sigma_2=-\omega_c\gamma$. 
The right hand side (r.h.s.) of Eq.~(\ref{eq:proof1}) crosses the y-axis at $\mu^2+|\alpha|\,\omega_c(1-\rho)$.
Since $\rho\,\in\,[-1,\,1]$ the y-axis is always crossed at positive values if  $\mu\neq 0$ and there is at least one intersection with $\sigma>0$ independently of the branches for $\pm\rho$.
There could however be an intersection at zero if $\mu=0$ and $\rho=1$.
This occurs if $|\zeta|\cos(-\Psi)=1$, which is only true for $|\zeta|=1$ and $\Psi=2\pi\,n$ $(n\in\mathbb{Z})$, i.e., related to a global phase shift as previously discussed. 
This concludes the proof and hence, for $\alpha<0$ there always exists at least one $\sigma>0$.
Hence, the direction $\alpha<0\rightarrow\sigma>0$ of the stability criterion in \cite{Earl2003} holds in the presence of inertia.

Now, we show that in regimes where the perturbation response dynamics are overdamped, i.e., $\mu=0$, the stability criterion 
holds also for second order phase models.
Hence, \textit{for $\mu=0$ and if $\alpha>0$ there can only be solutions with $\sigma<0$}. 
Let us consider the contrary, for $\mu=0$ and if $\alpha>0$ there always exists at least one solution with $\sigma\geq 0$. 
In that case we would have $\mu=0$, $\sigma =|\sigma|$ and $\alpha=|\alpha|$.
Using these expressions in Eq.~(\ref{eq:Squared_Real_Imag_Parts}) we find
\begin{equation}
  \frac{(|\sigma|^2+\omega_c\gamma|\sigma|+|\alpha|\omega_c)^2}{(|\alpha|\omega_c)^2}=|\zeta|^2e^{-2|\sigma|\tau}.
  \label{eq:Squared_gamma_0}
\end{equation}
The r.h.s. is always in $[0,1]$ due to $|\zeta|\,\in\,[0,1]$ as shown before using Gershgorin's circle theorem and $e^{-2|\sigma|\tau}\in[0,1]$ for $\sigma \geq 0$.
For $\sigma>0$ the l.h.s. is always larger than $1$ 
which contradicts $|\zeta|^2e^{-2|\sigma|\tau}\leq 1$.
%
The l.h.s. can only be equal to $1$ for $\sigma=0$ which leads to equality with the r.h.s for $\zeta=\pm 1$ only.
For $\zeta=-1$ while $\lambda=0$ we know that $\alpha$ has to be zero, see Eq.~(\ref{eq:characteristic_equation}), which contradicts the assumptions.
The case $\zeta=1$ relates to a global phase shift and is not considered as discussed before.
As the contrary can never be fulfilled, the original proposition is always true.

Using the same graphical procedure as before in Fig.~\ref{fig:graphic_solutions} (left)
we now ask for $\mu\neq0$, whether \textit{if $\alpha > 0$, there always exists at least one $\sigma\geq 0$}.
Setting $\alpha=|\alpha|$ in Eq.~(\ref{eq:real_part}) it can be shown that the proposition cannot always be fulfilled when studying
the r.h.s. for $\sigma=0$.
If the asymptotic value of the r.h.s. is $\mu^2 - |\alpha|\,\omega_c<0$ and we consider the branch for $\rho>0$, then if $\mu^2 - |\alpha|\,\omega_c(1-|\rho| )<0$ only solutions at $\sigma<0$ can exist.
For $\rho<0$ there cannot be a solution for $\sigma\geq0$ and a solution at $\sigma<0$ cannot be guaranteed.
Hence, the proposition cannot always be fulfilled, bifurcations can occur when $\alpha>0$.

We proceed to derive sufficient and necessary criteria that identify parameter regimes where the in- and anti-phase synchronized states in Eq.~(\ref{eq:SyncStates}) are unstable when $\alpha>0$ and $\mu \neq 0$.
Studying the properties of Eqs.~(\ref{eq:real_and_imaginary_parts})
at $\sigma=0$ and taking into account their asymptotic properties, stability criteria that connect $\tilde{\mu}=\mu(\sigma=0)$ and the parameters are obtained.
We ask when at least one solution with $\sigma>0$ exists.
Rearranging Eq.~(\ref{eq:imaginary_part}) and setting $\hat{\rho}=|\zeta|\sin(\mu\tau-\Psi)$ we find $2\sigma\mu = -\omega_c\left(\mu\gamma+|\alpha|\hat{\rho}\,e^{-\sigma\tau}\right)$.
Four cases $\{\pm\tilde{\mu},\,\pm\hat{\rho}\}$ need to be distinguished.
The cases for $\tilde{\mu}=-|\tilde{\mu}|$ are shown in Fig.~\ref{fig:graphic_solutions} (middle).
%
Using the asymptotic property of the r.h.s. of Eq.~(\ref{eq:imaginary_part}) reveals that for the case $\alpha>0$ and $\hat{\rho}<0$ there cannot be an intersection at $\sigma\geq 0$.
For $\hat{\rho}>0$ and $|\tilde{\mu}| > |\alpha||\hat{\rho}|/\gamma$ no intersection at $\sigma \geq 0$ can exist.
In the other cases when $\tilde{\mu}=|\tilde{\mu}|$ one finds that for $\hat{\rho}>0$ there 
cannot be an intersection at $\sigma\geq 0$.
For $\hat{\rho}<0$ no solutions at $\sigma\geq 0$ can exist if $|\tilde{\mu}| > |\alpha||\hat{\rho}|/\gamma$.
The proof to conclude necessity has the same structure.
Our criteria are in agreement with abstract mathematical results obtained for real $\zeta$ \cite{Pontryagin1955}.

\textit{Applying these criteria to study linear stability:}---
The criteria derived in the last section can only be meaningfully applied if the $\tilde{\mu}$ are known.
We calculate $\tilde{\mu}$ at the bifurcation,
i.e., at the critical point $\sigma=0$. 
Hence, side-bands at $\Omega\pm \tilde{\mu}$ arise in the power spectrum \cite{Pollakis2014}.
With these $\tilde{\mu}$ linear stability can be analyzed as a function of the network topology, the interaction strength, the damping coefficient, the time delay and the inertial parameter.
Moreover, we obtain a condition that predicts how the bifurcation
can be prevented based only on the physical parameters of the system.
Setting $\sigma=0$ while $\alpha=|\alpha|$, Eq.~(\ref{eq:Squared_Real_Imag_Parts}) after rearranging becomes 
\begin{equation}
  \tilde{\mu}^4+\tilde{\mu}^2(\gamma^2\omega_c^2-2|\alpha|\omega_c)+(|\alpha|\omega_c)^2(1-|\zeta|^2)=0.
  \label{eq:polynomial_sigma_0}
\end{equation}
Demanding $\tilde{\mu}\,\in\,\mathbb{R}$ while $\alpha>0$, a condition where synchronized states in Eq.~(\ref{eq:SyncStates}) are stable is obtained:
\begin{equation}
 \frac{\omega_c\gamma^2}{2|\alpha|}>1-\sqrt{1-|\zeta_0|^2}.
   \label{eq:condition_noInstab}
\end{equation}
Here $\zeta_0$ denotes the eigenvalue with the largest magnitude.
This result can be combined with the analysis of the criteria derived in the previous section using the solutions $\tilde{\mu}$ obtained from Eq.~(\ref{eq:polynomial_sigma_0}).
For the case of $\zeta=-1$ ($\Psi=\pi$), e.g., the case for $N=2$ mutually coupled oscillators, the r.h.s and l.h.s. of Eq.~(\ref{eq:imaginary_part}) become zero when $\tilde{\mu}=0$.
Such $\tilde{\mu}$ are actually solutions to Eq.~(\ref{eq:polynomial_sigma_0}) in this special case.
This would imply that any type of perturbation response is a valid solution.
Consulting Eq.~(\ref{eq:real_part}) for such $\tilde{\mu}=0$ when $\zeta=-1$ it becomes clear that additional information is necessary to infer whether or not the 
bifurcation has occurred. 
We need to plot Eq.~(\ref{eq:imaginary_part}) for $\sigma\geq 0$ in the $\mu-f(\mu)$ plane and ask when additional solutions $\mu\neq 0$ can arise that lead to 
bifurcations, see Fig.~\ref{fig:graphic_solutions} (right).
%
From studying the slopes at $\mu=0$ and the smallest $|\sigma|=0$ we find that if $\gamma\geq |\alpha|\omega_c$ no additional solutions with $\sigma\geq0$ can exist and hence the state is linearly stable.
An overview on how to apply the criteria is provided in the Supplementary Materials.

\textit{Parameter space plots analyzing linear stability:}---
All parameter space plots share the same color code.
When plots cover parameter space where multiple synchronized states are stable, the stability of the one with the largest frequency $\Omega=\{\Omega_i\}_{i\in\mathbb{N}}^\textrm{max}$ is plotted.
Python scripts that implement these criteria are available online \cite{GitHubRepo}.
These can also solve Eq.~(\ref{eq:characteristic_equation}) numerically for validation purposes, see examples provided in the Supplementary Materials.

Cyan denotes regimes where $\alpha<0$ and the in- or anti-phase synchronized state, see Eq.~(\ref{eq:SyncStates}), is unstable.
States with different constant phase relations exist in these regimes and can be stable if $\alpha>0$.
Purple regimes denote where in- or anti-phase synchronized states are unstable due to inert system behavior when $\alpha>0$. 
They are qualitatively different from parameter regimes where synchronized states that satisfy Eq.~(\ref{eq:SyncStates}) become unstable when $\alpha<0$.
At their onset they are characterized by time-dependent frequencies and highly correlated periodic dynamics.
Hence, synchronization in a wider sense is not necessarily lost after the Hopf bifurcation.
There are indications that these systems undergo a route to chaos via subsequent period-doubling bifurcations as, e.g., the time delay is increased \cite{Punetha2019}.
\begin{figure}[pt!]
\includegraphics[width=3.0in]{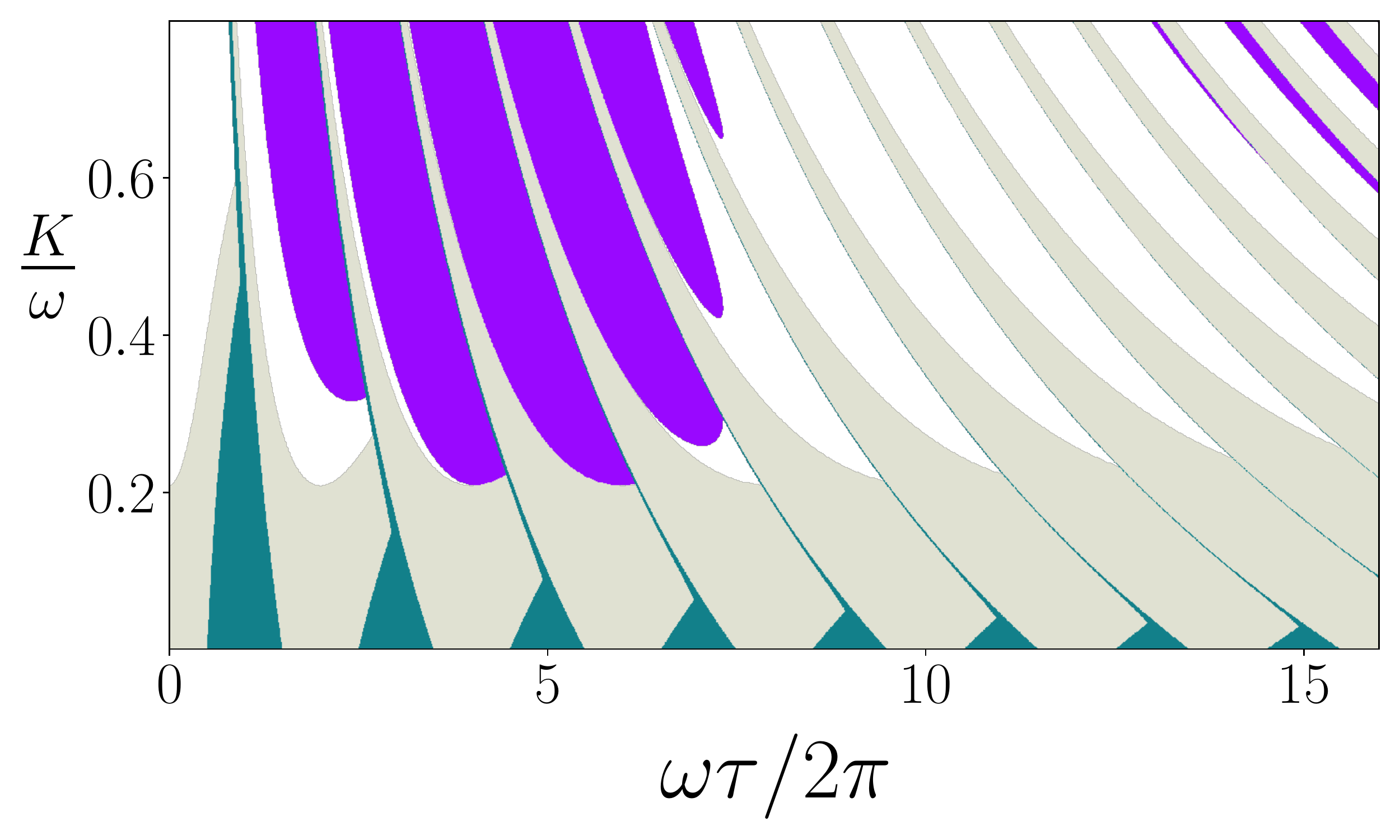} 
\caption{$K$ vs $\tau$ parameter space for $3\times3$ identical oscillators mutually coupled to their nearest neighbors on a $2d$ square grid with periodic boundary conditions. Parameters are $\omega=2\pi$\,radHz, $v=1$, $\omega_c=0.028\pi$\,radHz, $\gamma=2$, $\zeta\in\{-0.5,\,0.25\}$. Unstable regimes shown in cyan/purple, stable in white/grey.
}
\label{fig:tau_vs_K}
\end{figure}

Stable synchronized states are shown in grey and white.
Grey specifies where the condition in Eq.~(\ref{eq:condition_noInstab}) is fulfilled for the $\zeta_0$ with the largest magnitude.
In general, the synchronized states of Eq.~(\ref{eq:SyncStates}) tend to become unstable as the oscillators become increasingly inert ($\omega_c \ll \omega$)
and the purple regimes cover larger areas in Fig.~\ref{fig:tau_vs_K}.
Since $\omega_c$ plays an important role in suppressing higher order frequency contributions in real systems, it cannot be increased freely.
Above some critical $\omega_c$ the presence of, e.g., intermodulation products which are not described in Eq.~(\ref{eq:model_2nd}), can also lead to time-dependent frequencies.
Our results can guide towards optimal parameter choices for applications, e.g., in coupled electronic oscillators.

\begin{figure}[pt!]
    \includegraphics[width=3.0in]{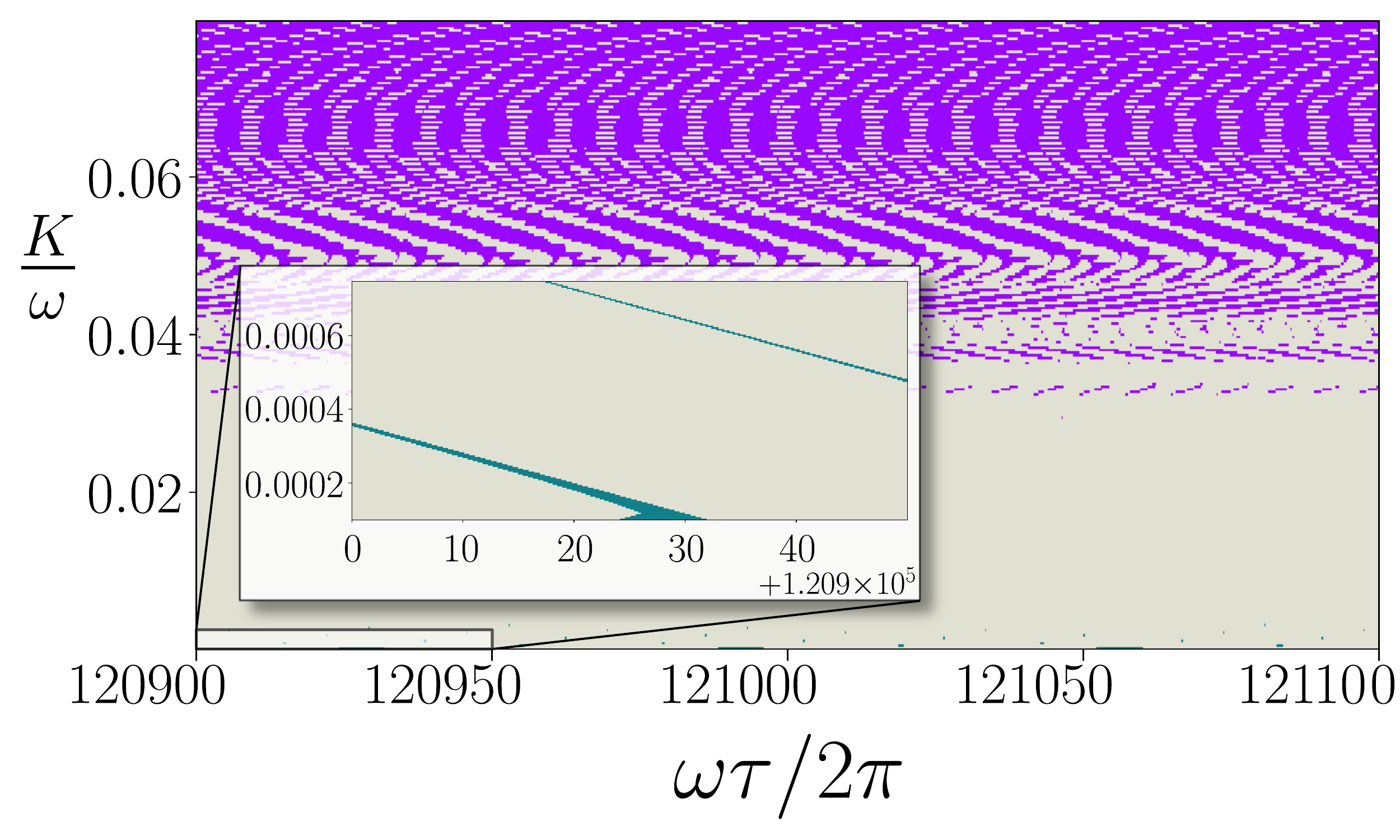}
\caption{$K$ vs $\tau$ parameter space for $3\times3$ identical oscillators with nearest neighbor coupling on a $2d$ square grid with open boundary conditions. 
Parameters are $\omega=2\pi$\,radHz, $\gamma=1$, $v=64$, $\omega_c=0.0007\pi$\,radHz, and $\zeta\in\{-1,\,-0.5,\,0.5\}$. 
Cyan structures in inset are not visible in main due to resolution.
}
\label{fig:tau_vs_K_large}
\end{figure}

\begin{figure}[pb!]
    \includegraphics[width=3.0in]{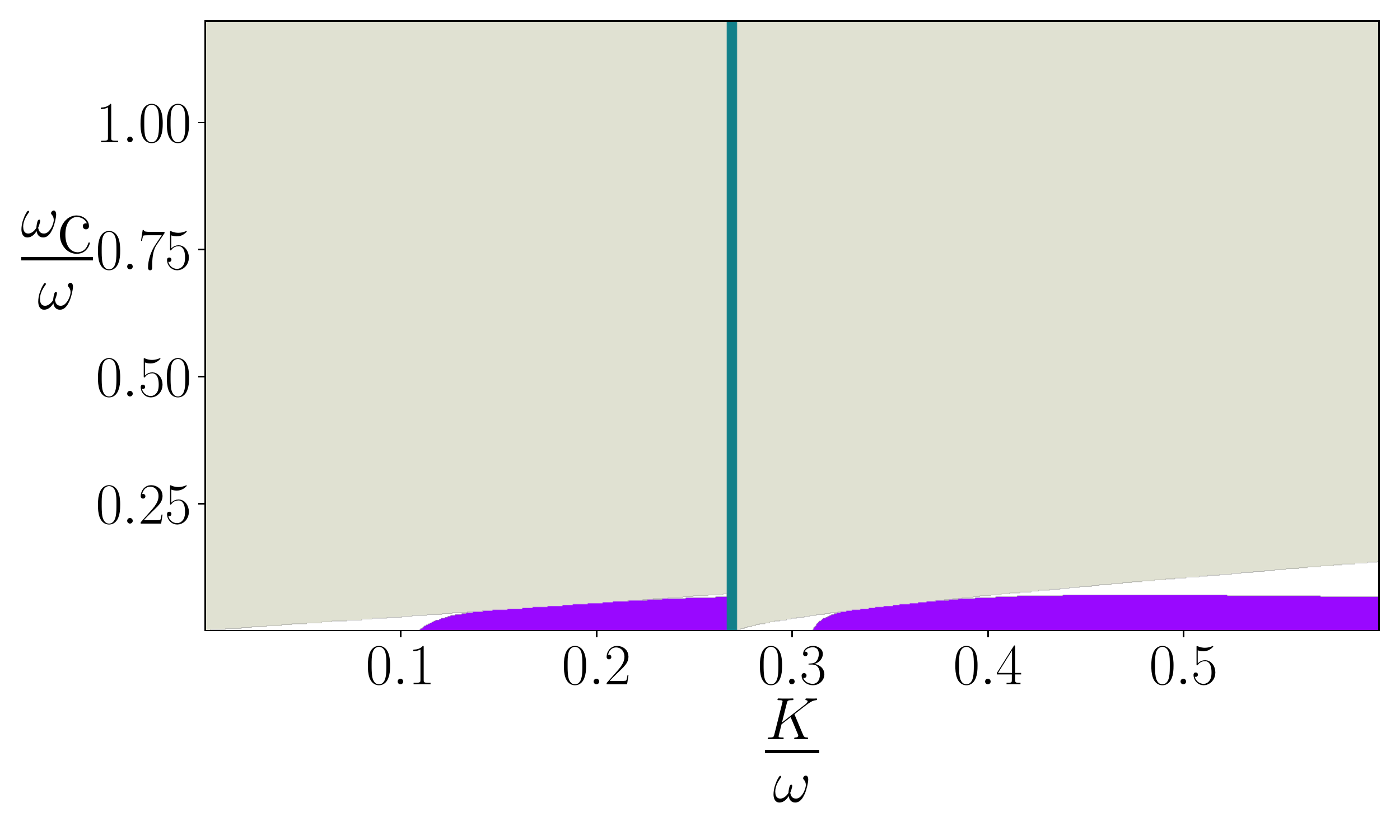} 
\caption{$\omega_c$ vs $K$ parameter space.  
Parameters, network topology and color code as in Fig.~\ref{fig:tau_vs_K}, except $\gamma=1$ and $\tau=0.65$\,s.
}
\label{fig:K_vs_wc}
\end{figure}

\textit{The physics of synchronization for large delays:}---
Networks of mutually coupled electronic oscillators, so called phase-locked loops \cite{Goldman2007}, are candidates for enabling new technologies, such as satellite independent terrestrial navigation and to provide orchestration to complex spatially distributed systems. 
Their function relies on a robust clock signal distribution.
%
Given operational frequencies up to the THz regime, spatial extensions of a few hundreds of meters imply time delays that are $3-6$ orders of magnitude larger than the oscillation period.
In these cases synchronization can only be stable for adequately divided cross-coupling frequencies \cite{Hoyer2021}.
This also requires to sufficiently decrease $\omega_c$, i.e., making the oscillators more inert.
Otherwise side-bands will appear in the frequency spectrum that may lead to, e.g., cross-channel interference \cite{Toscano2008}.
In consequence, the loop gains $\alpha$ have to be tuned sufficiently small to prevent violation of the condition in Eq.~(\ref{eq:condition_noInstab}) as $\omega_c$ is decreased.
Our results also show that $\omega_c$ can be optimized beyond this condition, see white spaces in Fig.~\ref{fig:K_vs_wc}.
Another challenge is the large number of synchronized states that can exist simultaneously.
As a result, it becomes difficult to determine stability numerically or in simulations.
Using the criteria we derived, the stability at arbitrary time delays can now be obtained.
We find that synchronization is feasible even when time delays span thousands of the oscillators' periods, see Fig.~\ref{fig:tau_vs_K_large}.
The condition in Eq.~(\ref{eq:condition_noInstab}) involves a periodic dependence on the time delay via $\alpha$.
This suggests that fine-tuning the delay can enable stable synchronization at very large time delay.
In real systems this may be limited by signal degradation during sending and dynamic noise.
Note also, that for $N$ all to all coupled oscillators $\zeta_0=(N-1)^{-1}$.
Hence, the stable regime guaranteed by condition Eq.~(\ref{eq:condition_noInstab}) increases with $N$.

\begin{figure}[pt!]
    \includegraphics[width=3.0in]{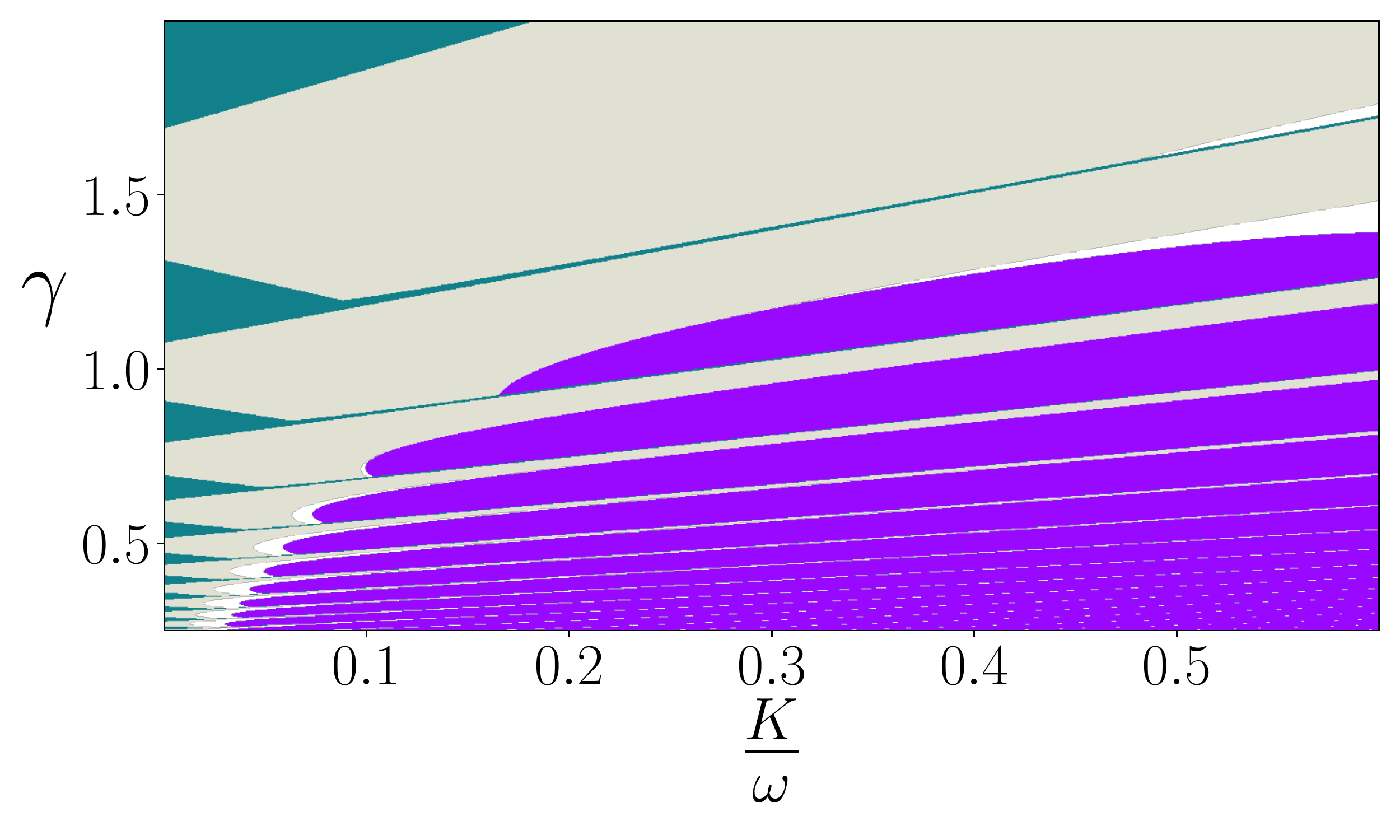}
\caption{$\gamma$ vs $K$ parameter space.
Parameters, network topology and color code as in Fig.~\ref{fig:tau_vs_K}, except $\gamma=1$, $\omega_c=0.4\pi$ and $\tau=2.95$\,s. 
}
\label{fig:K_vs_fric}
\end{figure}

\begin{figure}[pt!]
    \includegraphics[width=3.0in]{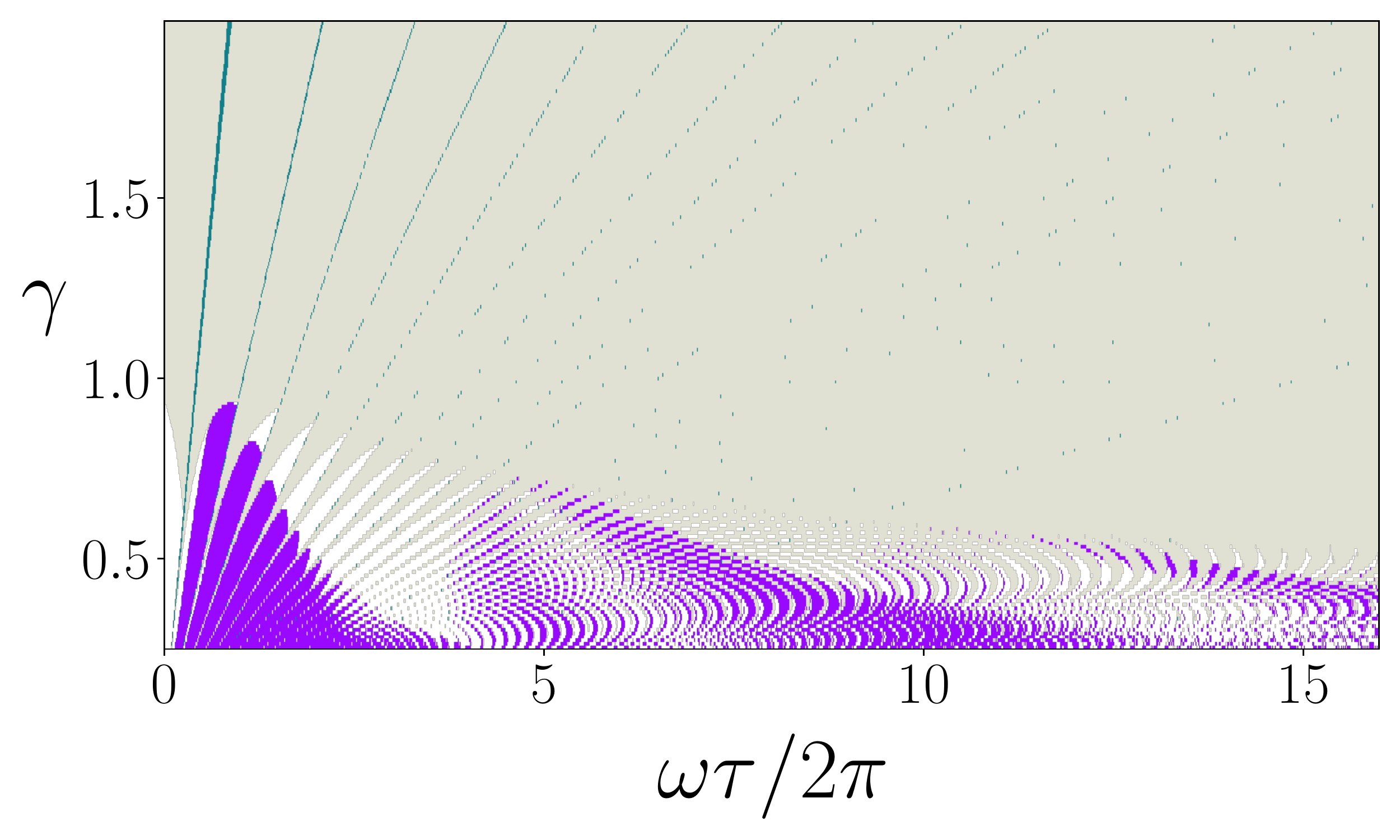}
\caption{$\gamma$ vs $\tau$ parameter space.
Parameters, network topology and color code as in Fig.~\ref{fig:tau_vs_K}, except $\omega_c=0.4\pi$\,radHz and $K=1.3\pi$\,radHz.}
\label{fig:tau_vs_fric}
\end{figure}

\textit{Damping coefficient rescales delay and frequency:}---
$\gamma$ relates to, e.g., gains in electronic oscillators, a friction in mechanical or damping coefficent in power grid systems.
Substituting $\Omega^*=\gamma\Omega$ and $\tau^*=\tau/\gamma$ in Eq.~(\ref{eq:SyncStates}) reveals that $\gamma$ acts as a rescaling of the time delay and frequency of synchronized states.
The relation between time delay and period of the oscillations changes,  observe the repetitive cyan-colored structures where $\alpha<0$ in Fig.~\ref{fig:K_vs_fric}. 
Decreasing $\gamma$ below one increases the frequency $\Omega$ of a synchronized state.
For constant $\omega_c$ that changes the ratio $\omega_c/\Omega$ and can trigger inertia-induced
bifurcations, see Fig.~\ref{fig:tau_vs_fric}.

\textit{Discussion and conclusions:}---
We derived general stability criteria for in- and anti-phase synchronized states in systems of delay-coupled oscillators with inertia for the first time. 
With their help, we identify parameter regimes where system's with inert oscillator response can excite additional frequencies or lead to chaotic dynamics.
In a synchronized state with $\{\Omega,\,\beta\}$ constant in time,
$\alpha$ denotes the change of the oscillators interaction terms with respect to a small perturbation.
For $\alpha<0$ the interaction between the oscillators becomes repelling.
In this case another stable synchronized state exists for which $\alpha>0$ and the coupling is attractive.
As long as perturbation responses are overdamped no bifurcations occur.
That changes for underdamped response.
Then, our criteria and condition Eq.~(\ref{eq:condition_noInstab}) reveal how the interplay between the oscillators' parameters and those of the network lead to bifurcations.
A specific example is how the cutoff frequency $\omega_c$ of a filter limits the loop gain $\alpha$ in networks of electronic oscillators, see condition Eq.~(\ref{eq:condition_noInstab}).
In the presence of dynamic noise, this analysis can be carried out within the Fokker-Planck formalism \cite{Yeung1999, Acebron2000} and is subject to ongoing work. 
 
Our analysis can be applied to various fields as the response of natural systems usually is inert \cite{Lindsey1985,Trees2005,Dewenter2015}.
It is especially helpful when the numerical solution of the characteristic Eq.~(\ref{eq:characteristic_equation}) or simulations become infeasible.
For applications, our results enable fast identification of the parameter regimes where synchronized states with constant phase-differences are stable.
%
This will improve the architecture design process of, e.g., networks of electronic oscillators \cite{Lindsey1985, Wetzel2021}.
Furthermore it can enable real time control algorithms for on-the-fly optimization of such complex systems, e.g., when topology or time delays change over time.
We show that stable mutual synchronization is feasible at large time delay.
This makes it a candidate for the next generation self-organized clocking signal distribution layers \cite{Patent2014SyncNet}.
It is relevant for, e.g., precise localisation using micro-satellites or terrestrial beacons, sensoring and time distribution, high precision physical measurements in spatially distributed systems such as very long baseline interferometry and gravitational wave detection \cite{Loschmidt2009, Punturo2010, Schuh2012, Cappelletti2021}.
\begin{acknowledgments}
We would like to pay our regards to C. Zheng, P. Pnigouras, H. Kantz, D. Schmidt, D. J\"{o}rg, A. Pollakis, G. Fettweis, and F. J\"{u}licher for inspiring discussions. 
The work on this topic was initiated within the Cluster of Excellence Center for Advancing Electronics Dresden.
This work was supported by the Federal Ministry of Education and Research (BMBF) under the reference number 03VP06431.
D. Prousalis and L. Wetzel contributed equally to this work. 
\end{acknowledgments}

\bibliography{References}

\end{document}



\title{Supplementary Material: Synchronization in the presence of time delays and inertia: Stability criteria}

\author{Dimitrios Prousalis}%
 \email{dprou@pks.mpg.de}
\author{Lucas Wetzel}
 \email{lwetzel@pks.mpg.de}
\affiliation{%
Max Planck Institute for the Physics of Complex Systems, Dresden, Germany
}%

\date{\today}
\maketitle
Here we provide technical details on the calculations and theorems used in the main text.
We present a numerical verification of the conditions and criteria presented using parameter plots obtained numerically from the characteristic equation.
Furthermore the criteria are summarized in a table for convenient application.
The parameters of all plots in the main and Supplementary Material are shown in Table \ref{tab:parameters_main} and \ref{tab:parameters_supp}.
We note, that examples of time-series of the solutions that bifurcate from in- and anti-phase synchronized states in the purple marked regimes can be found in Nirmal et al., Supplementary Materials \cite{Punetha2019}.

\subsection{How to apply the critera and conditions}

In a first step the properties of the synchronized states whose linear stability is to be studied have to be computed using
\begin{equation}
 \gamma\,\Omega=\omega+K\,h\left(-\frac{\Omega\tau+\beta}{v}\right),
 \label{eq:SyncStates}
\end{equation}
This yields the global frequencies $\Omega$ and phase-differences $\beta$ of the existing synchronized states.
From this, the parameter
\begin{equation}
 \alpha=\frac{K}{v} \, h'\left(\frac{-\Omega \tau+\beta}{v}\right),
 \label{eq:alpha}
\end{equation}
can be obtained. 
In a next step the eigenvalues $\zeta$ of the normalized adjacency matrix need to be calculated.
Then, solving Eq. (10) we can calculate $\tilde{\mu}$ from:
\begin{equation}
\tilde{\mu}=\pm\sqrt{-\frac{A}{2} \pm \frac{A}{2} \sqrt{1-\frac{4B}{A^2}}  }\, ,
\label{eq:solutions_gamma}
\end{equation}
where $A(\tau,K,\omega_c)=\gamma^2\omega_c^2-2|\alpha|\omega_c$ and $B(\tau, K,\omega_c)=(|\alpha|\omega_c)^2(1-|\zeta|^2)$\, .
Demanding that $\tilde{\mu}\,\in\,\mathbb{R}$, conditions where $\sigma$ changes its sign while $\alpha>0$ are obtained:
\begin{equation}
 \frac{\omega_c\gamma^2}{2|\alpha|}>1-\sqrt{1-|\zeta_0|^2}.
  \label{eq:condition_noInstab}
\end{equation}  
Then condition (\ref{eq:condition_noInstab}) can be used to understand where synchronized states must be stable. 
\begin{table}[!ht]
\begin{center}

 \begin{tabular}{|c||c||} 
 \hline
sufficient conditions & stability  \\ [0.5ex] 
 \hline\hline
 $\alpha<0$ & unstable\\ 
 \hline
 $\alpha>0$ and $\mu=0$ & stable\\
 \hline
 $\alpha>0$ and $1-\sqrt{1-|\zeta_0|^2}<\frac{\omega_c\gamma^2}{2|\alpha|}$ & stable\\
    \hline
\end{tabular}
\end{center}
\caption{Stability conditions}
\label{tab:stab_cons}
\end{table}
\begin{table}[!ht]
\begin{center}

 \begin{tabular}{|c||c||} 
 \hline
sufficient and necessary criteria & stability  \\ [0.5ex] 
 \hline\hline
 $\tilde{\mu}<0$ and  $\hat{\rho}<0$ & stable\\
  $\tilde{\mu}<0$  and  $\hat{\rho}>0$  and $|\tilde{\mu}| > |\alpha||\hat{\rho}|/\gamma$ & \\
  \hline
 $\tilde{\mu}>0$  and  $\hat{\rho}>0$& stable\\ 
 $\tilde{\mu}>0$  and  $\hat{\rho}>0$ and $|\tilde{\mu}| > |\alpha||\hat{\rho}|/\gamma$ &\\
 \hline
 $\zeta=-1$ and $\tilde{\mu}=0$  and $\gamma\geq |\alpha|\omega_c$ & stable\\
 \hline
\end{tabular}
\end{center}
\caption{Stability criteria for $\alpha>0$ and $\mu\neq0$, where $\hat{\rho}=|\zeta|\sin(\tilde{\mu}\tau-\Psi)$
}
\label{tab:stab_criteria}
\end{table}
If this condition is not fulfilled and $\tilde{\mu}\,\in\,\mathbb{R}$, the criteria in Table~\ref{tab:stab_criteria} identify where states with time-dependent frequencies have emerged from the ones with constant frequencies.
In Tables~\ref{tab:stab_cons},~\ref{tab:stab_criteria} we show all the criteria and conditions which need to be checked in order to study the linear stability of in- and anti-phase synchronized states with constant frequency. 

\subsection{Gershgorin theorem}

The Gershgorin theorem defines disks centered around each diagonal entry of a matrix $\mathbb{D}$ with radius given by the row sum over all non-diagonal entries \cite{Gerschgorin31}.
As there is no self-coupling all $d_{kk}=0\,\,\,\forall\,k$ and hence all Gershgorin discs are centered at the origin.
Hence $|\zeta|$ is bounded by one
\begin{equation}
 |\zeta|\leq\ \sum\limits_{l \neq k}\,|d_{kl}|=1.
 \label{eq:Gershgorin}
\end{equation}
Taking into account the definition $d_{kl}=c_{kl}/n_k$, the boundedness of the $\zeta$'s given by Gershgorin's theorem is tied to the identical coupling capacity of all oscillators
\begin{equation}
 \frac{1}{n_k}\sum\limits_{l \neq k}\,|c_{kl}|=1.
 \label{eq:coupCapacity}
\end{equation}
This can also be achieved when the oscillators weight different inputs with different weights, i.e.,
\begin{equation}
 \frac{1}{n_{kl}}\sum\limits_{l \neq k}\,|c_{kl}|=1.
 \label{eq:coupCapacity2}
\end{equation}
when the sum over all non-diagonal entries equals to one.

\subsection{Additional plots and numerical verification}

\begin{figure}[!pb]
    \includegraphics[width=1\linewidth]{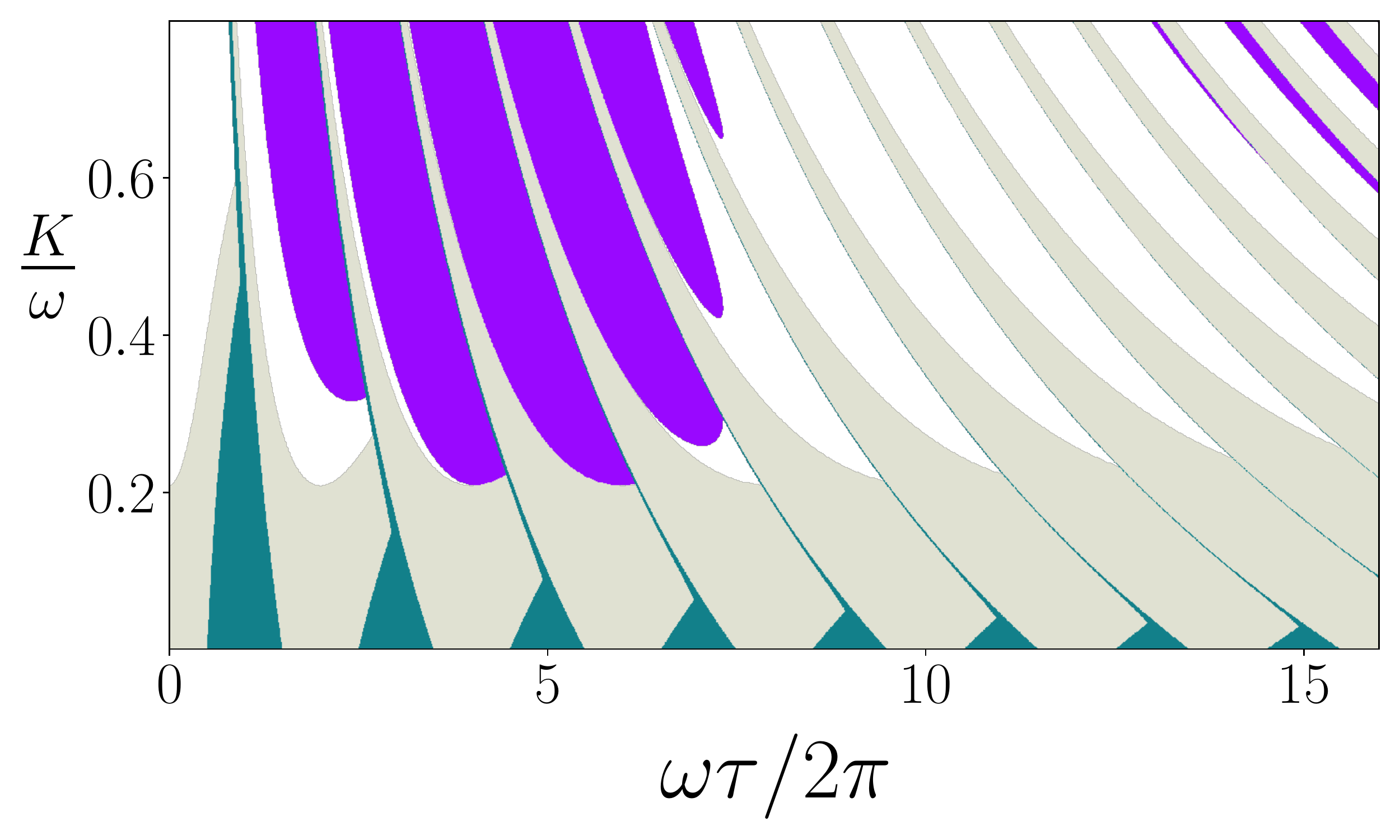}
     \caption{$K$ vs $\tau$ parameter space for $3\times3$ mutually delay-coupled oscillators on a $2$d square grid with periodic boundary conditions, obtained analytically via the criteria and condition. All other parameters listed in Table~\ref{tab:parameters_supp}.}
     \label{Supfig:Criteria_tau_vs_K_3x3}
\end{figure}

\begin{figure}[!hpt]
  \includegraphics[width=1\linewidth]{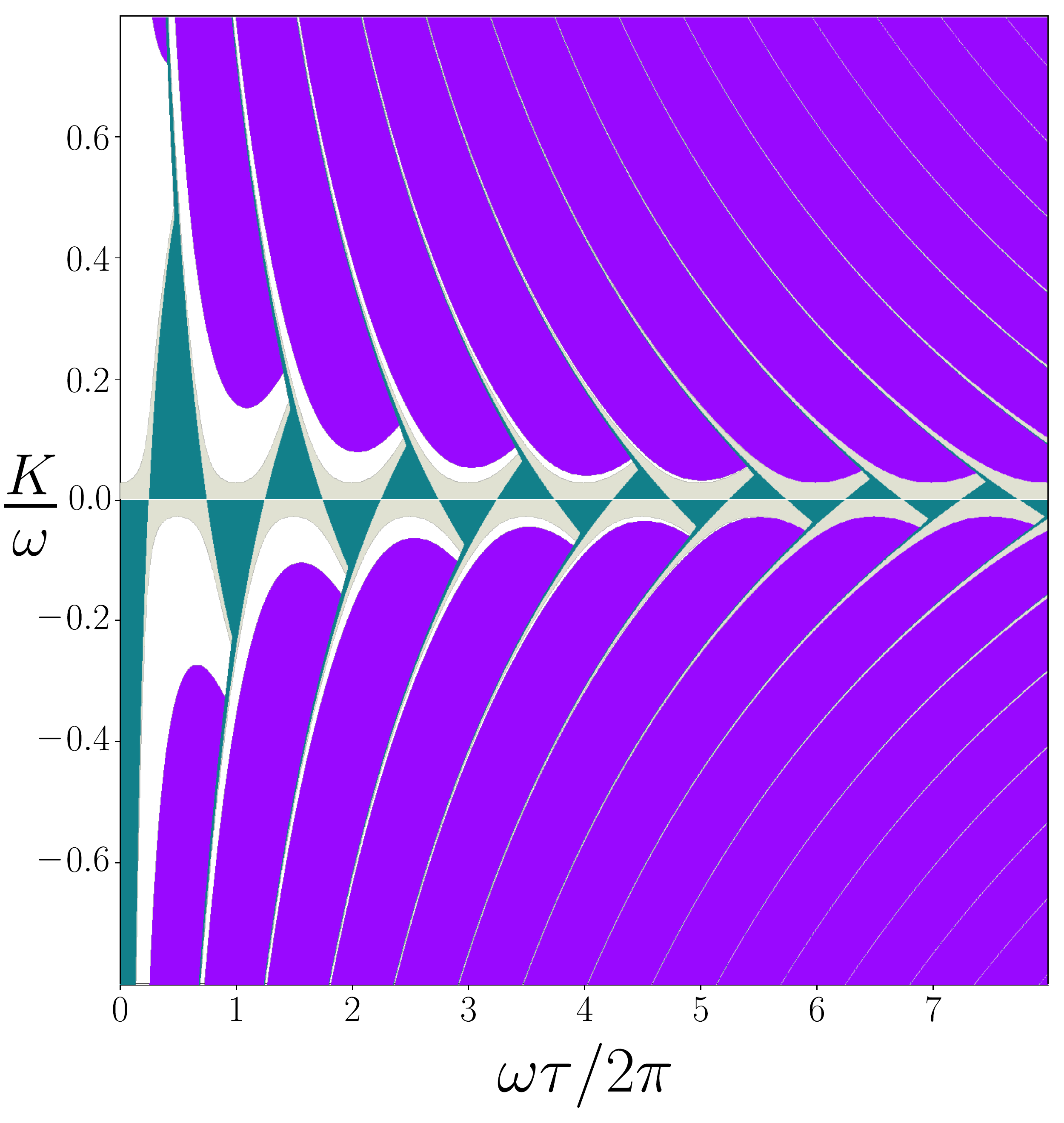}
   \caption{$K$ vs $\tau$ parameter space for $N=2$ mutually delay-coupled oscillators, obtained analytically via the criteria and condition. In- ($K\geq0$) and anti-phase ($K<0$) synchronized states are shown. All other parameters listed in Table~\ref{tab:parameters_supp}.}
   \label{Supfig:Criteria_tau_vs_K_N2_InphaseAntiphase}
\end{figure}

The parameter space plots, Figs.
\ref{Supfig:Criteria_tau_vs_K_3x3},
\ref{Supfig:Com_Criteria_tau_vs_K_N2},   \ref{Supfig:Criteria_tau_vs_K_N2_InphaseAntiphase},  share the same color code. 
Stable synchronized states are shown in grey and white.
Grey specifies where the condition in Eq.~(\ref{eq:condition_noInstab}) is fulfilled.
The white regimes relate to states which are stable, when $\alpha>0$, i.e., the criteria for the $\tilde{\mu}\in\mathbb{R}$ are fulfilled. 
\begin{figure}[!pt]
  \includegraphics[width=1\linewidth]{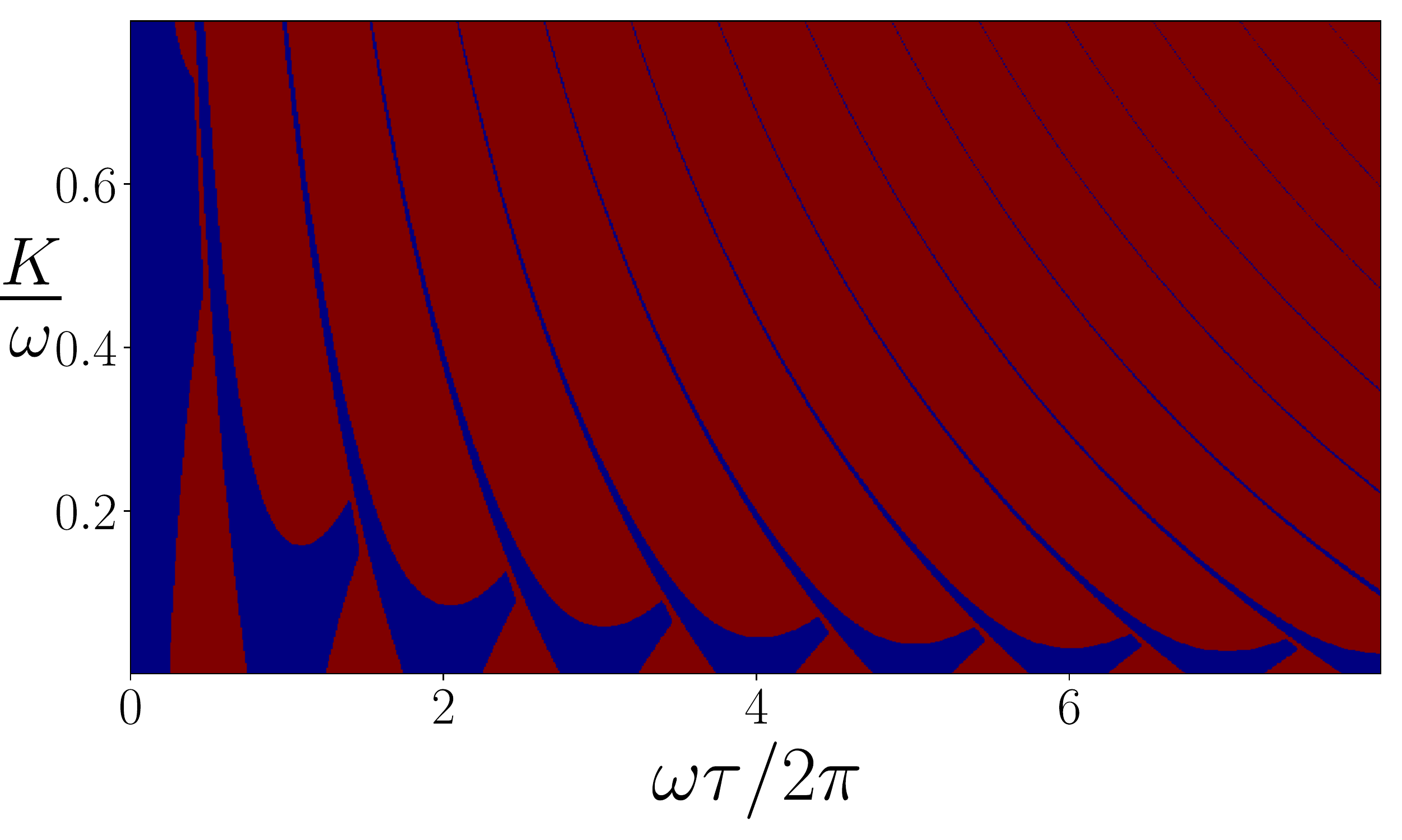}
  \caption{$K$ vs $\tau$ parameter space for $N=2$ mutually delay-coupled oscillators, obtained numerically. All other parameters listed in Table~\ref{tab:parameters_supp}.}
    \label{Supfig:ReLambda_in_tau_vs_K2}
\end{figure}
Unstable synchronized states are shown in cyan and purple.
Cyan denotes regimes where $\alpha<0$ and synchronized states are unstable. 
The purple regimes relate to unstable states when $\alpha>0$.
\subsubsection{Numerical verification of the criteria}
We verify the results obtained using the criteria and the conditions with numerical solutions to the characteristic equation.
The numerical results are obtained using Python scripts which are available online \cite{GitHubRepo}.
The parameters of the system under investigation can be set in the dictionaries of these scripts.
\begin{figure}[!pt]
 \includegraphics[width=1\linewidth]{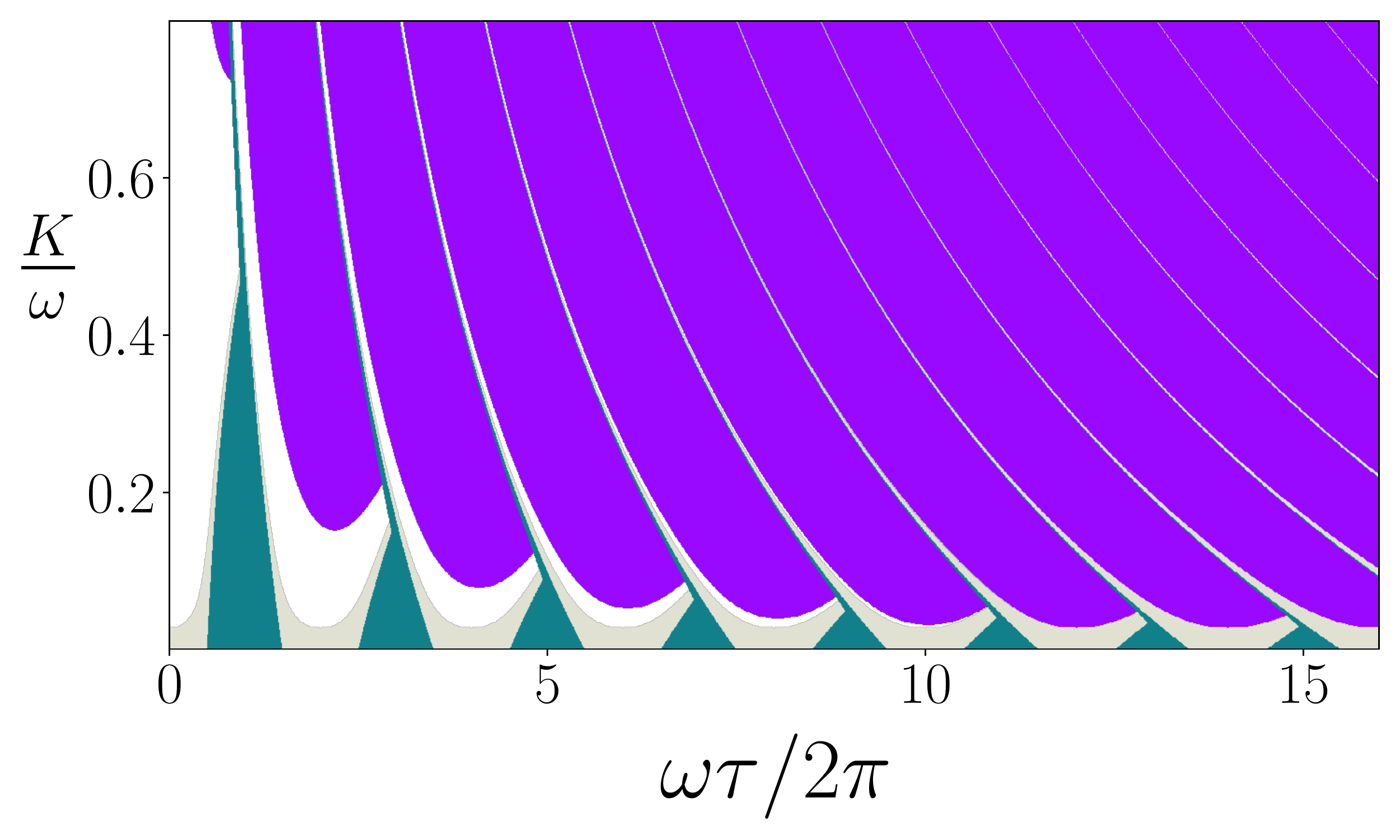}
 \caption{$K$ vs $\tau$ parameter space for $N=2$ mutually delay-coupled oscillators, obtained analytically via the criteria and condition. All other parameters listed in Table~\ref{tab:parameters_supp}.}
 \label{Supfig:Com_Criteria_tau_vs_K_N2}
\end{figure}
The numerically obtained parameter space plot in Figs.~\ref{Supfig:ReLambda_in_tau_vs_K2} has the following color code. 
%
%
In this case red denotes regimes where the real part $\sigma$ of the characteristic equation is positive and hence the synchronized states are unstable.
Blue denotes regimes where $\sigma$ is negative and in- and/or anti-phase synchronized states are stable. 
We choose this different color code for the numerical results to avoid confusion and clearly distinguish them from those obtained with the stability conditions and criteria.

It becomes apparent from the comparison of Figs.~\ref{Supfig:ReLambda_in_tau_vs_K2} and \ref{Supfig:Com_Criteria_tau_vs_K_N2} that the results obtained analytically using the criteria are in agreement with the numerical results.
The value of the perturbation response rate obtained from the numerical solution shows no qualitative difference for the white and grey regimes.
Hence, it is not represented by the color code in Fig.~\ref{Supfig:ReLambda_in_tau_vs_K2}.

\subsubsection{Parametric plots of different topologies}

Here we show parametric plots for different network topologies of mutually delay-coupled oscillators with inertia.
All other parameters have the same values, i.e., identical oscillator parameters and time delays. 

In Fig.~\ref{Supfig:Com_Criteria_tau_vs_K_N2}, the $K-\tau$ parameter space for a network of two mutually coupled oscillators is shown.
The nontrivial eigenvalue $\zeta$ of the adjacency matrix is $\zeta=-1$.
In Fig.~ \ref{Supfig:Criteria_tau_vs_K_3x3} the parameter space of a network of nine mutually delay-coupled oscillators in a $2$d lattice with nearest neighbor interactions and periodic boundary conditions is shown.
The nontrivial eigenvalues $\zeta$ of the adjacency matrix in this case are $\zeta=\{-0.5,0.25\}$.
It can be observed that linear stability is different for the two cases. 
The condition in Eq.~ (\ref{eq:condition_noInstab}) is fulfilled for different regimes, shown in grey in the figures.
When $\zeta=-1$ then grey covers the smallest regime, as can be understood from Eq.~(\ref{eq:condition_noInstab}).
For networks in which the oscillators are arranged in $1$d with open boundary conditions, e.g., a chain topology, the value $\zeta=-1$ is always in the set of $\zeta$'s \cite{Pollakis2014}. 
The criteria which are presented in Table~\ref{tab:stab_criteria} are also affected by the topology and hence the stability regimes in white and purple, are also different.
\begin{table}[!pb]
\begin{center}
 \begin{tabular}{||c |c |c |c |c |c | c |c||} 
 \hline
Fig & $\omega$ & $\omega_c$ &$\tau$& $\gamma$ & K & $\zeta$& v\\
 & radHz & radHz & sec& radHz& radHz&  &  \\ [0.5ex] 
 \hline\hline
 1,2,4& $2\pi$ & $0.028\pi$ & $-$& $2$& $-$& $-1$ & 1\\
 \hline
 3 & $2\pi$ & $0.028\pi$ & $-$& $2$& $-$& $\{-0.5,\,0.25\}$ & 1\\
 \hline
\end{tabular}
\end{center}
\caption{Parameters of plots in Supplemental Material}
\label{tab:parameters_supp}
\end{table}
\begin{table}[!pb]
\begin{center}
 \begin{tabular}{||c |c |c |c |c |c | c |c||} 
 \hline
Fig & $\omega$ & $\omega_c$ &$\tau$& $\gamma$ & K & $\zeta$& v\\
 & radHz & radHz & sec& radHz& radHz&  &  \\ [0.5ex] 
 \hline\hline
 2& $2\pi$& $0.028\pi$ & $-$ & $2$& $-$ & $\{-0.5,\,0.25\}$& 1\\ 
 \hline
 3 & $2\pi$ & $0.0007\pi$& $-$& 1 & $-$& $\{-1, -0.5,\,0.25\}$ & 1\\
 \hline
 4 &$2\pi$ & $-$ & 0.65 & 1& $-$ &$\{-0.5,\,0.25\}$ & 1 \\
 \hline
 5& $2\pi$ & $0.4\pi$ & 2.95& $-$& $-$& $\{-0.5,\,0.25\}$ & 1\\
 \hline
 6 & $2\pi$& $0.4\pi$ & $-$& $-$ & $1.3\pi$& $\{-0.5,\,0.25\}$ & 1\\  [1ex] 
 \hline
\end{tabular}
\end{center}
\caption{Parameter values of the in plots in main text}
\label{tab:parameters_main}
\end{table}

\subsection{Parametric plot of in- and anti-phase synchronized states}
In Fig.~\ref{Supfig:Criteria_tau_vs_K_N2_InphaseAntiphase} we show the parametric $K$-$\tau$ plot for in- and anti-phase synchronized states in a system of two  mutually delay-coupled oscillators with inertia.
The regime for $K>0$ relates to the in-phase synchronized states and for $K<0$ to the anti-phase synchronized states.
Observe how the cyan regimes alternate  between in- and anti-phase synchronized state as the time delay increases.

\bibliography{References}